\documentclass[a4paper,11pt]{article}

 \usepackage{amsfonts,amsthm,amssymb}
 \usepackage{latexsym,color, epsf}

\def\picill#1by#2(#3)
{\vbox to #2 {\hrule width #1 height 0pt depth 0pt
\vfill\epsffile{#3}}}

\newcommand{\eq}{\begin{equation}}
\newcommand{\en}{\end{equation}}
\newcommand{\eqa}{\begin{eqnarray}}
\newcommand{\ena}{\end{eqnarray}}

\begin{document}

\setlength{\unitlength}{1mm}

\thispagestyle{empty}


\begin{flushright}
\end{flushright}


 \begin{center}
  { \bf   Teleportation, Braid Group and Temperley--Lieb Algebra
  \\[2mm]}

  \vspace{.5cm}

Yong Zhang${}^{ab}$\footnote{yzhang@nankai.edu.cn},

 ${}^a$  Theoretical Physics Division, Chern Institute of Mathematics \\
  Nankai University, Tianjin 300071, P. R. China\\[0.1cm]

  ${}^b$ Institute of Theoretical Physics, Chinese Academy of Sciences\\
 P. O. Box 2735, Beijing 100080, P. R. China \\[0.2cm]

\end{center}

\vspace{0.2cm}

\begin{center}
\parbox{13cm}{
\centerline{\small  \bf Abstract}  \noindent\\

We explore algebraic and topological structures underlying the
quantum teleportation phenomena by applying the braid group and
Temperley--Lieb algebra. We realize the braid teleportation
configuration, teleportation swapping and virtual braid
representation in the standard description of the teleportation. We
devise diagrammatic rules for quantum circuits involving maximally
entangled states and apply them to three sorts of descriptions of
the teleportation: the transfer operator, quantum measurements and
characteristic equations, and further propose the Temperley--Lieb
algebra under local unitary transformations to be a mathematical
structure underlying the teleportation. We compare our
diagrammatical approach with two known recipes to the quantum
information flow: the teleportation topology and strongly compact
closed category, in order to explain our diagrammatic rules to be a
natural diagrammatic language for the teleportation.

}

\end{center}

\vspace{.5cm}

\begin{tabbing}

\end{tabbing}


\newpage

\section{Introduction}

 Quantum entanglements \cite{werner1} play key roles in quantum
 information phenomena \cite{nielsen} and they are widely exploited in quantum
 algorithms \cite{shor, grover}, quantum cryptography
 \cite{bb84, ekert} and quantum teleportation \cite{bennett}.
 On the other hand, topological entanglements \cite{kauffman1}
 represent topological configurations like links or knots which are closures of
 braids. There are natural similarities between quantum entanglements
 and topological entanglements. As a unitary braid has a power of
 detecting knots or links, it often can transform  a separate quantum
 state into an entangled one. Hence a nontrivial unitary braid
 representation can be
 identified with a universal quantum gate \cite{BB, dye}. Recently,
 a series of papers have been published on the application of knot
 theory to quantum information, see \cite{kauffman2, molin1, molin2}
 for universal quantum gates and unitary solutions of the
 Yang--Baxter equation \cite{yang, baxter}; see \cite{kauffman3,
 kauffman4, kauffman5} for quantum topology and quantum computation;
 see \cite{kauffman6, kauffman7} for quantum entanglements and
 topological entanglements.

 Especially, Kauffman's work on the teleportation topology \cite{kauffman2, kauffman8}
 motivates our tour of revisiting in a diagrammatic approach all tight
 teleportation and dense coding schemes in Werner's paper \cite{werner2}.
 Under the project of setting up a bridge between knot theory and
 quantum information, the joint paper with Kauffman and Werner \cite{ZKW} explores
 topological and algebraic structures underlying multipartite entanglements
 by recognizing the Werner state as a rational solution of the
 Yang--Baxter equation and the isotropic state as a braid
 representation, and constructing a representation of the
 Temperley--Lieb algebra in terms of maximally entangled states,
 while the present paper focuses on the problem of how to study the teleportation
 from the viewpoints of the braid group and Temperley--Lieb algebra \cite{lieb}.

 The teleportation is a kind of quantum information protocol transporting
 a unknown quantum state. To describe it in a unified mathematical
formalism, we have to integrate standard quantum mechanics with
classical features since outcomes of quantum measurements are sent
to Bob from Alice via classical channels and then Bob carries out a
required unitary operation. The one approach has been proposed by
Abramsky and Coecke in recent research. It applies the category
theory to quantum information protocols and describes the quantum
information flow by strongly compact closed categories, see
\cite{coecke2, coecke3} for abstract physical traces; see
\cite{coecke1, coecke4} for the quantum information flow; see
\cite{coecke5, coecke6} for a categorical description of quantum
protocols; see \cite{coecke7, coecke8, coecke9} for diagrammatic
quantum mechanics and see \cite{duncan} for quantum logic.

As Abramsky and Coecke suggest \cite{abramsky}, we also expect a
powerful mathematical framework to describe quantum information
phenomena in a unified framework. We  believe in the existence of
beautiful mathematical structures underlying entanglement and
teleportation such as the braid group and Temperley--Lieb algebra
which are well known to the community of knot theory for a long
time. They not only simplify complicated algebraic calculations in
an intuitive  manner but also catch essential points of quantum
phenomena and exhibit them in a natural style\footnote{If one
accepts the validity of quantum mechanics which is justified by
enormous amount of experiments, then one should not state that
quantum teleportation, a valid result in this framework, would be a
mystery in any sense, no matter how counterintuitive it is.
Moreover, teleportation would be not at all surprising in the
framework of classical mechanics, where even cloning is possible. }.

 A maximally entangled bipartite state is found to form a representation of
 the Temperley--Lieb algebra. In view of
 the diagrammatic representation for the Temperley--Lieb algebra, we
 are inspired to deal with quantum information protocols involving maximally
 entangled states in a diagrammatic approach. We think that diagrams catch
 essential points from the global view so that they can express complicated
 algebraic objects in a simpler way. We represent maximally entangled
 vectors by cups or caps because they are widely exploited in topics
 including the Temperley--Lieb algebra, braid representations, knot theory
 and statistics mechanics \cite{kauffman1}.

Section 2 revisits the quantum teleportation from the viewpoints of
the braid group and virtual braid group \cite{kauffman9, kauffman10,
kauffman11,kamada}. The transformation matrix
 between the Bell states and product bases is found to form a
 braid representation and this stimulates us to propose the braid
 teleportation configuration together with the teleportation
 swapping and explain it with the crossed measurement
 \cite{vaidman1, vaidman, vaidman2}. Also, the virtual mixed relation
 for defining the virtual braid group is found to be a formulation of the
 teleportation equation.

Section 3 devises diagrammatical rules for quantum  information
protocols involving maximally entangled states, projective
measurements and local unitary transformations. Various properties
of maximally entangled states are collected and these guide us to
set up diagrammatical rules for assigning a definite diagram to a
given algebraic expression. Three types of descriptions for the
quantum teleportation phenomena: the transfer operator
\cite{preskill}, quantum measurements \cite{vaidman1, vaidman,
vaidman2} and characteristic equations \cite{werner2}, are
respectively revisited in our diagrammatical approach.

Section 4 proposes the Temperley--Lieb algebra under local unitary
transformations to be a suitable mathematical structure underlying
the quantum teleportation phenomena. The connections between the
diagrammatical representation for the Temperley--Lieb algebra and
our diagrammatical approach are made as clear as possible.  The
teleportation configuration is recognized as a fundamental
ingredient for defining the Brauer algebra \cite{brauer}, and it can
be performed in terms of swap gates and Bell measurements.

 Section 5 sketches two known diagrammatical approaches to the
 quantum information flow: Kauffman's teleportation topology
 \cite{kauffman2, kauffman8} and the categorical theory mainly
 considered by Abramsky and Coecke, which are compared with our
 diagrammatical approach in order to stress conceptual
  differences in both physics and mathematics among them and
  propose our diagrammatical rules to present a natural
  diagrammatic language for the teleportation phenomena.

Section 6 is on concluding remarks and outlooks. Our next steps in
this promising research are discussed in a brief way.

\section{Teleportation, braid group and virtual braid group}

Based on the teleportation equation in terms of the Bell matrix for
the standard description of the quantum teleportation phenomena, we
realize the braid configuration $(b^{-1}\otimes Id)(Id\otimes b)$
together with the teleportation swapping $(P\otimes Id)(Id\otimes
P)$, and explain it via the concept of the crossed measurement
\cite{vaidman1,vaidman, vaidman2}. We also study the teleportation
in terms of a virtual braid representation.

\subsection{Teleportation equation in terms of Bell matrix}

In terms of product bases $|ij\rangle$, $i,j=0,1$, the four mutually
orthogonal Bell states  have the forms, \eqa
  & & |\phi^+\rangle=\frac 1 {\sqrt 2} (|00\rangle+|11\rangle),
  \qquad |\phi^-\rangle=\frac 1 {\sqrt 2} (|00\rangle-|11\rangle),
  \nonumber\\
  & & |\psi^+\rangle=\frac 1 {\sqrt 2} (|01\rangle+|10\rangle),
  \qquad |\psi^-\rangle=\frac 1 {\sqrt 2} (|01\rangle-|10\rangle),
\ena which are transformed to each other under local unitary
 transformations,
 \eqa
 \label{local}
 & & |\phi^{-} \rangle=(1\!\! 1_2\otimes \sigma_3) |\phi^+\rangle =
  (\sigma_3 \otimes 1\!\! 1_2 ) |\phi^+\rangle, \nonumber \\
 &&  |\psi^+\rangle =(1\!\! 1_2\otimes \sigma_1) |\phi^+\rangle
  =(\sigma_1 \otimes 1\!\! 1_2) |\phi^+\rangle,
 \nonumber\\
 & & |\psi^-\rangle =( 1\!\! 1_2\otimes -i \sigma_2) |\phi^+\rangle
  =(i \sigma_2\otimes 1\!\! 1_2) |\phi^+\rangle,
 \ena
where $1\!\! 1_2$ denotes a $2\times 2$ unit matrix, so $1\!\! 1_d$
for a $d\times d$ unit matrix, and the Pauli matrices $\sigma_1$,
$\sigma_2$ and $\sigma_3$ have the conventional formalisms.

The teleportation is a quantum information protocol of sending a
message $|\psi\rangle_C$ from Charlie to Bob under the help of Alice
\footnote{Teleportation is usually considered as a protocol between
two parties: Alice and Bob, and it requires classical communication.
The third party, Charlie (who prepares the teleported quantum
state), has to send it to Alice directly (who shall perform a
measurement on it in order to have it teleported). }. Alice, who
shares a maximally entangled state $|\phi^+\rangle_{AB}$ with Bob,
performs an entangling measurement on the composite system between
Charlie and her and then informs results of her measurements to Bob,
who will know what Charlie wants to pass onto him according to a
protocol between Alice and him. Note the following calculation, also
see \cite{preskill}, \eqa \label{tele}
 & & |\psi\rangle_C|\phi^+\rangle_{AB} \equiv \frac 1 {\sqrt 2}
 ( a|0\rangle + b |1\rangle)_C (|00\rangle
 +|11\rangle)_{AB}   \\
  &=& \frac 1 2 ( |\phi^+\rangle_{CA}|\psi\rangle_B +
     |\phi^-\rangle_{CA} \sigma_3|\psi\rangle_B
     +  |\psi^+\rangle_{CA} \sigma_1|\psi\rangle_B
     +|\psi^-\rangle_{CA} (-i\sigma_2)|\psi\rangle_B ) \nonumber
\ena which is called the teleportation equation  and tells how to
teleport a qubit $|\psi\rangle_C$ from Charlie to Bob. When Alice
detects the Bell state $|\phi^+\rangle_{CA}$ and informs Bob about
that through a classical channel, Bob will know that he has a
quantum state $|\psi\rangle_B$. Similarly, when Alice gets the Bell
states $|\phi^-\rangle_{CA}$ or $|\psi^+\rangle_{CA}$ or
$|\psi^-\rangle_{CA}$, Bob will apply the local unitary
transformations $\sigma_3$ or $\sigma_1$ or $i\sigma_2$ on the
quantum state that he has  in order to obtain $|\psi\rangle_B$.

We introduce the Bell matrix \cite{kauffman2, molin1,dye} and denote
it by $B=(B_{ij,\,lm})$, $i,j,l,m=0,1$. The Bell matrix and
 its inverse or transpose  are given by  \eq \label{bell}
 B=\frac 1 {\sqrt 2}\left(
 \begin{array}{cccc}
 1 & 0 & 0 & 1 \\
 0 & 1 & -1 & 0 \\
 0 & 1 & 1 & 0 \\
 -1 & 0 & 0 & 1 \\
 \end{array} \right), \qquad  B^{-1}=B^{T}=\frac 1 {\sqrt 2}\left(
 \begin{array}{cccc}
 1 & 0 & 0 & -1 \\
 0 & 1 & 1 & 0 \\
 0 & -1 & 1 & 0 \\
 1 & 0 & 0 & 1 \\
 \end{array} \right).
\en It has an exponential formalism given by \eq
 B=e^{i\frac {\pi} 4 (\sigma_1\otimes \sigma_2)}=\cos{\frac {\pi}
 4}+ i \sin \frac {\pi} 4 (\sigma_1\otimes \sigma_2)
\en with the following interesting properties: \eq
B^2=i\sigma_1\otimes \sigma_2, \qquad B^4=-1\!\! 1_4, \qquad
B^8=1\!\! 1_4, \qquad
 B=\frac 1 {\sqrt 2} (1\!\! 1_4 + B^2). \en

In terms of the Bell matrix and product bases, the Bell states can
be generated in the formalism, \eqa
 & & |\phi^+\rangle=B|11\rangle, \qquad  |\phi^-\rangle=B|00\rangle,
 \nonumber\\
  & & |\psi^+\rangle=B|01\rangle, \qquad
  |\psi^-\rangle=-B|10\rangle
 \ena
where the Bell operator acts on product bases in the way, \eq
 B|ij\rangle =\sum_{k,l=0}^{1} |kl\rangle B_{kl,ij}=
 \sum_{k,l=0}^{1} |kl\rangle B^T_{ij,kl},
\en and hence the teleportation equation (\ref{tele}) can be
rewritten into a new formalism,
 \eq
 \label{btele}
 (1\!\! 1_2 \otimes B)(|\psi\rangle\otimes |11\rangle)_{CAB}
 \equiv (B\otimes 1\!\! 1_2)
 ({\vec v}^T \otimes \frac 1 2 {\vec\sigma}_{11} |\psi\rangle
 )_{CAB},
 \en
where the vectors $\vec{\sigma}_{11}$ and  $\vec{v}$ are convenient
notations, their transposes given by \eq
 \vec{\sigma}_{11}^{T}\equiv(\sigma_3, \sigma_1, i\sigma_2, 1\!\! 1_2),
  \qquad {\vec v}^T \equiv (|00\rangle,|01\rangle,|10\rangle,|11\rangle) \en
  and the calculation of ${\vec v}^T \otimes {\vec\sigma}_{11}$ follows
  a rule: \eq {\vec v}^T \otimes {\vec\sigma}_{11} |\psi\rangle \equiv
|00\rangle\otimes \sigma_3 |\psi\rangle
 + |01\rangle\otimes \sigma_1 |\psi\rangle
 + |10\rangle\otimes i\sigma_2 |\psi\rangle
 +  |11\rangle\otimes  |\psi\rangle.   \en

The remaining three teleportation equations are derived in the same
way by applying local unitary transformations among the Bell states
(\ref{local}) to the teleportation equation (\ref{btele}). As a
maximally entangled state shared by Alice and Bob is
$|\phi^-\rangle_{AB}$, the teleportation equation has the form
 \eqa
 & & |\psi\rangle_C|\phi^-\rangle_{AB}= (1\!\! 1_2 \otimes B)
  (|\psi\rangle\otimes |00\rangle)_{CAB}
 \nonumber\\
 &=&|\psi\rangle_C\otimes (1\!\! 1_2\otimes \sigma_3)|\phi^+\rangle_{AB}
 = (B\otimes 1\!\! 1_2)({\vec v}^T \otimes \frac 1 2 \sigma_3{\vec\sigma}_{11}
 |\psi\rangle)_{CAB}
 \ena
where the local unitary transformation $1\!\! 1_2\otimes1\!\!
1_2\otimes \sigma_3$ commutes with $B\otimes 1\!\! 1_2$.
 Similarly, the other two teleportation equations are obtained in the
 following, \eqa & &
  (B^{-1}\otimes 1\!\! 1_2) (1\!\! 1_2 \otimes B)(|\psi\rangle\otimes
|01\rangle)_{CAB}=({\vec v}^T \otimes \frac 1 2
\sigma_1{\vec\sigma}_{11} |\psi\rangle)_{CAB}, \nonumber\\
 & & (B^{-1}\otimes 1\!\! 1_2)(1\!\! 1_2 \otimes B)(|\psi\rangle\otimes -|10\rangle)_{CAB}
  =({\vec v}^T \otimes -\frac 1 2 i \sigma_2{\vec\sigma}_{11} |\psi\rangle )_{CAB}. \ena
It is obvious that the matrix configuration $(B^{-1}\otimes 1\!\!
1_2)(1\!\! 1_2 \otimes B)$ plays a key role in the above
teleportation equations in terms of the Bell matrix.

 \subsection{Braid teleportation configuration $(b^{-1} \otimes Id)( Id \otimes b)$}

\begin{figure}
\begin{center}
\epsfxsize=10.5cm \epsffile{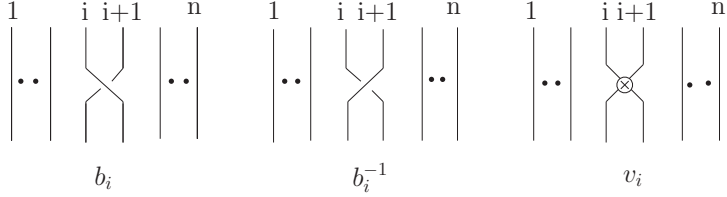} \caption{Braid generators
$b_i$, $b_i^{-1}$ and virtual braid generator $v_i$. } \label{fig1}
\end{center}
\end{figure}

 We sketch the definitions for the braid group and virtual braid group
and verify the Bell matrix $B$ to form a braid representation and
virtual braid representation. We propose the braid teleportation
configuration $(b^{-1} \otimes Id)( Id \otimes b)$ with the
teleportation swapping as its special example, and explain it in
terms of the crossed measurement \cite{vaidman1}.

A braid representation $b$-matrix is a $d \times d $ matrix acting
on $ V\otimes V$ where $V$ is a $d$-dimensional complex vector
space. The symbol $b_i$ denotes a braid $b$  acting on the tensor
product $V_i\otimes V_{i+1}$, see Figure 1 where $b_i$ is described
by a under crossing and its inverse $b_i^{-1}$ is represented by an
over crossing. The classical braid group $B_{n}$ is generated by
braids $b_1, b_2$, $\cdots, b_{n-1}$ satisfying the braid relation,
see Figure 2,
  \eqa \label{bgr}
   b_{i} b_{j} &=& b_{j} b_{i}, \qquad  j \neq i \pm 1, \nonumber\\
  b_{i}  b_{i+1} b_{i} &=& b_{i+1} b_{i} b_{i+1},
  \qquad i=1, \cdots, n-2.
 \ena

\begin{figure}
\begin{center}
\epsfxsize=11.5cm \epsffile{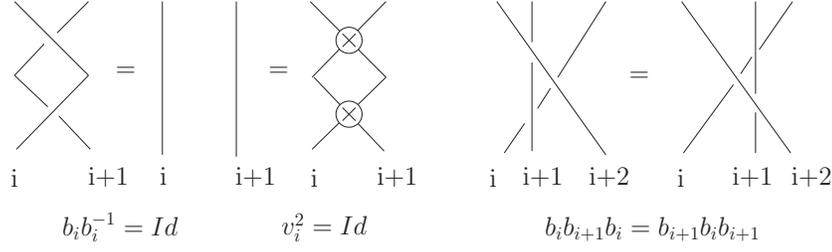} \caption{Identity
and the braid relation.} \label{fig2}
\end{center}
\end{figure}

 The virtual braid group $VB_n$ \cite{kauffman9, kauffman10, kauffman11, kamada}
 is an extension of the classical braid group $B_n$ by the symmetric group $S_n$.
 It has both braids $b_i$ and virtual crossings $v_i$ which are defined by the
 virtual crossing relation, \eqa \label{vbgr1}
 v_i^2 &=& 1\!\! 1, \qquad  v_i v_{i+1} v_i = v_{i+1} v_i v_{i+1},
 \qquad i=1,\cdots, n-2,
 \nonumber\\
 v_i v_j &=& v_j v_i, \qquad j \neq i \pm 1,
\ena a presentation of the symmetric group $S_n$, and the virtual
mixed relation: \eqa \label{vbgr2}  b_i v_j &=& v_j b_i,
\qquad   j \neq i \pm 1, \nonumber\\
 b_{i+1}v_{i}v_{i+1} &=& v_{i}v_{i+1} b_{i}, \qquad i=1,\cdots, n-2.
\ena

See Figure 1-2. A virtual crossing $v_i$ is represented by two
crossing arcs with a small circle placed around a crossing point. In
virtual crossings, we do not distinguish between under and over
crossings but which are described respectively in the classical knot
theory. The identity $Id$ is represented by
 parallel vertical straight lines without any crossings.

We verify the Bell matrix to satisfy the braid relation (\ref{bgr}).
On its right handside of (\ref{bgr}), after a little algebra we have
 \eq
  (1\!\! 1_2\otimes B)(B\otimes 1\!\! 1_2)(1\!\! 1_2\otimes B)
   = \frac 1 {\sqrt{2}} (1\!\! 1_2\otimes B^2
      + B^2\otimes 1\!\! 1_2)
 \en
and on its left handside we can derive the same result. We now prove
the Bell matrix to satisfy the virtual mixed relation (\ref{vbgr2})
as the permutation matrix $P$ is chosen as a virtual crossing,
\eq \label{permutation}  P=\left(\begin{array}{cccc} 1 & 0 & 0 & 0\\
 0 & 0 & 1 & 0\\
 0 & 1 & 0 & 0 \\
 0 & 0 & 0 & 1
\end{array} \right), \qquad P|ij\rangle=|ji\rangle, \qquad i,j=0,1.
\en On the left handside of (\ref{vbgr2}), we derive \eq (1\!\! 1_2
\otimes B)(P\otimes 1\!\!
 1_2)(1\!\! 1_2 \otimes P)(|i\rangle \otimes |j\rangle \otimes|k\rangle)
 = (1\!\! 1_2 \otimes B)(|k\rangle\otimes |ij\rangle)
 \en
which can be also obtained via the right handside of (\ref{vbgr2}).

\begin{figure}
\begin{center}
\epsfxsize=12.cm \epsffile{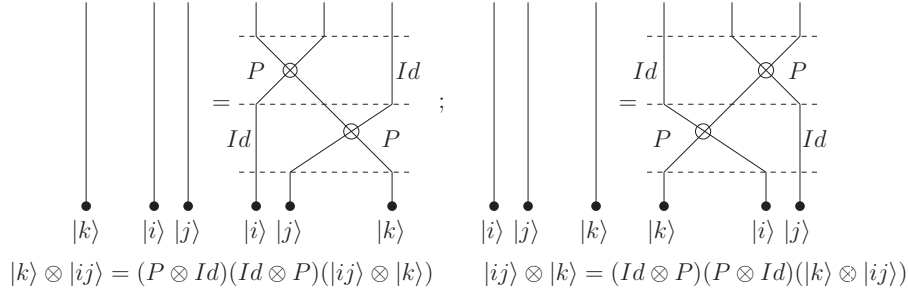} \caption{Teleportation
swapping with permutation $P$ as virtual crossing.} \label{fig3}
\end{center}
\end{figure}

 In view of the fact that the braid configuration $(B^{-1}\otimes 1\!\!
1_2)(1\!\! 1_2 \otimes B)$ is the most important element in the
above teleportation equation (\ref{btele}), we propose the concept
of the braid teleportation configuration $(b^{-1} \otimes Id)( Id
\otimes b)$.  Since a braid is a kind of generalization of
  permutation, we call $(P\otimes Id)(Id \otimes P)$ or $(Id \otimes P)(P\otimes Id)$
  as the teleportation swapping, which satisfy \eqa & &
  |k\rangle \otimes |ij\rangle
 =(P\otimes Id)(Id\otimes P) (|ij\rangle \otimes |k\rangle),
 \nonumber\\
 & & |ij\rangle\otimes |k\rangle
 =(Id\otimes P)(P\otimes Id) (|k\rangle \otimes |ij\rangle).
 \ena
See Figure 3, the permutation $P$ represented by a virtual crossing
with a small circle at the crossing point. In the crossed
measurement \cite{vaidman1}, a braid or crossing acts as a device of
measurement which is non-local in both space and time. In Figure 4,
the two lines of a crossing $b$ represent two observable operations:
the first relating the measurement at the space-time point $(x_1,
t_1)$ to that at the other point $(x_2, t_2)$ and the second one
relating the measurement at $(x_1,t_2)$ to that at $(x_2,t_1)$. The
crossed measurement $(Id\otimes b)$ plays a role of sending a qubit
from Charlie to Alice with a possible local unitary transformation.
Similarly, the crossed measurement $(b^{-1}\otimes Id)$ transfers a
qubit from Alice to Bob with a possible local unitary
transformation.

\begin{figure}
\begin{center}
\epsfxsize=10.5cm \epsffile{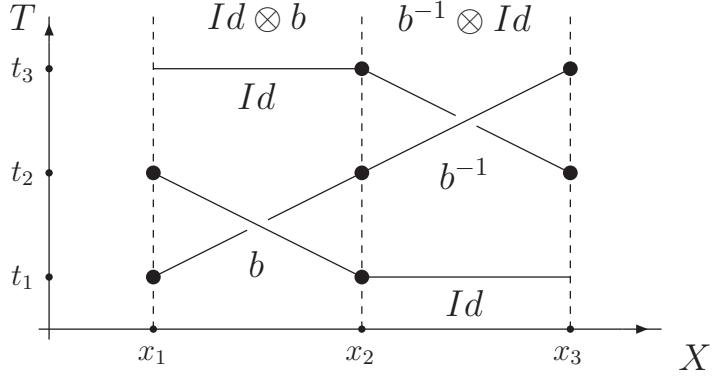} \caption{Braid
teleportation configuration and crossed measurement.} \label{fig4}
\end{center}
\end{figure}

\subsection{Teleportation and virtual braid group}

We describe the teleportation in the framework of the virtual braid
group: The braid relation (\ref{bgr}) builds a connection between
topological entanglements and quantum entanglements, while the
virtual mixed relation (\ref{vbgr2}) is a sort of reformulation of
the teleportation equation (\ref{tele}). A nontrivial unitary braid
detecting knots or links can be identified with a universal quantum
gate transforming a separate state into an entangled one, see
\cite{kauffman2, molin1, molin2}. In the following, we make a
connection clear between the virtual mixed relation (\ref{vbgr2})
and the teleportation equation (\ref{tele}).

In terms of the Bell matrix and teleportation swapping, the left
handside of the teleportation equation (\ref{tele}) has a
 form,
 \eq
 |\psi\rangle_C\otimes |\phi^+\rangle_{AB}=
  (1\!\! 1_2\otimes B)(P\otimes 1\!\! 1_2)(1\!\! 1_2\otimes P)
 (|11\rangle_{CA}\otimes |\psi\rangle_{B}),
 \en
while its right handside, $RHS$ leads to a formalism,
 \eq
 RHS= (1\!\! 1_2 \otimes P -1\!\! 1_2 \otimes \sigma_2 \otimes \sigma_2)
   (B\otimes 1\!\! 1 _2) (|11\rangle_{CA}\otimes |\psi\rangle_B)
 \en
where the permutation matrix $P$ (\ref{permutation}) is given by
 \eq
 \label{perm}
 P=\frac 1 2 (1\!\! 1_4 + \sigma_1\otimes \sigma_1
 +\sigma_2\otimes \sigma_2 +\sigma_3 \otimes
 \sigma_3),
 \en
and local unitary transformations (\ref{local}) among the Bell
states are used. Hence the teleportation equation (\ref{tele}) has a
new formulation given by \eq \label{vtele}
 |\psi\rangle_C\otimes|\phi^+\rangle_{AB} =(1\!\! 1_2\otimes
 P-1\!\! 1_2\otimes \sigma_2\otimes \sigma_2)(|\phi^+\rangle_{CA}\otimes
|\psi\rangle_{B}).
 \en

This equation (\ref{vtele}) is an equivalent realization of the
teleportation swapping on the state $|\phi^+\rangle_{CA}\otimes
|\psi\rangle_{B}$:
 \eq
(1\!\! 1_2\otimes
 P-1\!\! 1_2\otimes \sigma_2\otimes \sigma_2)(|\phi^+\rangle_{CA}\otimes
|\psi\rangle_{B})=(P\otimes 1\!\! 1_2)(1\!\! 1_2\otimes P)
(|\phi^+\rangle_{CA}\otimes |\psi\rangle_{B}), \en and it can be
regarded as a formulation of the virtual mixed relation
 (\ref{vbgr2}) on the state $|11\rangle_{CA}\otimes |\psi\rangle_{B}$,
 \eqa & & (1\!\! 1_2\otimes B)(P\otimes 1\!\! 1_2)(1\!\! 1_2\otimes P)
 (|11\rangle_{CA}\otimes |\psi\rangle_{B}) \nonumber\\
 & &  =(P\otimes 1\!\! 1_2)(1\!\! 1_2\otimes P)(B\otimes 1\!\! 1_2)
 (|11\rangle_{CA}\otimes |\psi\rangle_{B}).
\ena

Similarly, the teleportation equations for the Bell state
$|\phi^-\rangle_{AB}$, $|\psi^\pm\rangle_{AB}$ are respectively
obtained to be \eqa |\psi\rangle_C \otimes |\phi^-\rangle_{AB}
 &=& (1\!\! 1_2 \otimes P - 1\!\! 1_2 \otimes \sigma_1\otimes \sigma_1)
   (|\phi^-\rangle_{CA}\otimes |\psi\rangle_B), \nonumber\\
  |\psi\rangle_C \otimes |\psi^+\rangle_{AB} &=&
 (1\!\! 1_2 \otimes P - 1\!\! 1_2 \otimes \sigma_3\otimes \sigma_3)
   (|\psi^+\rangle_{CA}\otimes |\psi\rangle_B), \nonumber\\
  |\psi\rangle_C \otimes |\psi^-\rangle_{AB} &=&
 (1\!\! 1_2 \otimes P - 1\!\! 1_8)
   (|\psi^-\rangle_{CA}\otimes |\psi\rangle_B),
  \ena
in which local unitary transformations of
 $(1\!\! 1_2 \otimes P - 1\!\! 1_2 \otimes \sigma_2\otimes \sigma_2)$
 have been exploited. All of them can be identified with realizations of
 the virtual mixed relation (\ref{vbgr2}) or the teleportation
 swapping.

\section{Diagrammatical representations for teleportation}

We devise diagrammatical rules for describing maximally entangled
states in a diagrammatical approach and apply them to three typical
descriptions of the quantum teleportation phenomena: the transfer
operator, quantum measurements and characteristic equations.

\subsection{Notations for maximally entangled states}

 Maximally entangled states have various good algebraic properties and
 they play important roles in the quantum teleportation phenomena.
 Here we fix our notations for maximally entangled states.
The vectors $|e_i\rangle$ form a set of complete and orthogonal
bases for a  $d$-dimension Hilbert space $\cal H$, and the covectors
$\langle e_i|$ are chosen for its dual Hilbert space ${\cal
H}^\ast$, i.e., they satisfy \eq
 \sum_{i=0}^{d-1} |e_i\rangle \langle e_i |=1\!\! 1_d, \qquad
 \langle e_j| e_i\rangle =\delta_{ij}, \qquad i,j=0,1,\cdots d-1,
\en where $\delta_{ij}$ is the Kronecker symbol.

A  maximally entangled bipartite vector $|\Omega\rangle$ and its
 dual vector $\langle \Omega|$ have the forms
\eq
 |\Omega\rangle=\frac 1 {\sqrt{d}} \sum_{i=0}^{d-1} |e_i\otimes
 e_i\rangle,
 \qquad \langle \Omega | = \frac 1 {\sqrt{d}} \sum_{i=0}^{d-1}
 \langle e_i\otimes e_i |. \en
The local action of a bounded linear operator $M$ in the Hilbert
space $\cal H$ on $|\Omega\rangle$ satisfies \eq \label{matrix}
 (M\otimes 1\!\! 1_d) |\Omega\rangle =(1\!\! 1_d \otimes M^T) |\Omega\rangle,
  \qquad M_{ij} \equiv \langle e_i | M |e_j \rangle, \,\,\,
    M^T_{ij}=M_{ji}, \en and so it is
permitted to move the local action of the operator $M$ from the
Hilbert space to the other Hilbert space as $M$ acts on
$|\Omega\rangle$. A trace of two operators $M^\dag$ and $M^\prime$
can be represented by an inner product of two quantum vectors
$|\psi\rangle$ and $|\psi^\prime\rangle$,
 \eq  tr(M^\dag M^\prime)=d \, \langle \psi |\psi^\prime\rangle,
 \qquad   |\psi\rangle \equiv (M\otimes 1\!\! 1_d)|\Omega\rangle,\,\,\,
 |\psi^\prime\rangle \equiv (M^\prime\otimes 1\!\! 1_d)|\Omega\rangle,
 \en
 while an inner product with the action of an operator product $N_1\otimes N_2$
 is also a form of trace, \eq \langle \psi |
 N_1\otimes N_2 |\psi^\prime\rangle =\frac 1 d tr(M^\dag N_1 M^\prime
 N_2^T). \en

The transfer operator $T_{BC}$ sending a quantum state from Charlie
to Bob, \eq \label{transfer} T_{BC} \equiv
 \sum_{i=0}^{d-1} |e_i\rangle_B\,\, {}_C\langle e_i |, \qquad
 T_{BC} |\psi\rangle_C  = |\psi\rangle_B,
\en is recognized to be an inner product between maximally entangled
vectors $|\Phi(U) \rangle_{CA}$ and $|\Phi(V^T) \rangle_{AB}$
defined by local unitary actions of $U$ and $V^T$ on
$|\Omega\rangle$, i.e.,
 \eqa
 \label{transfer22}
 & & {}_{CA} \langle \Phi(U)| \Phi(V^T) \rangle_{AB} \equiv
 {}_{CA} \langle \Omega | ( U^\dag \otimes 1\!\! 1_d  ) ( V^T \otimes 1\!\! 1_d )
 |\Omega \rangle_{AB} \nonumber\\
 & & =  {}_{CAB} \langle \Omega\otimes 1\!\! 1_d |
  ( U^\dag \otimes 1\!\! 1_d \otimes 1\!\! 1_d )
   (1\!\! 1_d \otimes V^T \otimes 1\!\! 1_d )
 |1\!\! 1_d \otimes \Omega \rangle_{CAB} \nonumber\\
 & & = \frac 1 d (VU^\dag)_B \,\, T_{BC}
 \ena
which has a special case of $U=V$ given by \eq
 \frac 1 d T_{BC} = {}_{CA} \langle \Phi(U)| \Phi(U^T)\rangle_{AB}
  = {}_{CA}\langle \Omega | \Omega\rangle_{AB}.
\en

 A maximally entangled vector $|\Omega_n\rangle$ is a local unitary
 transformation
  of $|\Omega\rangle$, i.e., $|\Omega_n\rangle =(U_n \otimes 1\!\! 1_d)  |\Omega\rangle$,
  and the set of local unitary operators $U_n$ satisfies the
  orthogonal relation $tr(U_n^\dag U_m)=d\, \delta_{nm}$,
  which leads to the fundamental properties of $|\Omega_n\rangle$, \eq \label{ortho}
 \langle \Omega_n |\Omega_m \rangle =\delta_{nm}, \qquad \sum_{n=1}^{d^2}
  |\Omega_n\rangle \langle \Omega_n|=1\!\! 1_d,
  \qquad n,m=1,\cdots, d^2.
 \en We introduce the symbol $\omega_n$ to denote a maximally
 entangled state $|\Omega_n\rangle \langle \Omega_n|$ and especially
 use the symbol $\omega$ to represent $|\Omega\rangle \langle \Omega|$, i.e.,
 \eq
 \label{orthon}
 \omega\equiv|\Omega\rangle \langle \Omega|, \qquad \omega_n\equiv|\Omega_n\rangle \langle
 \Omega_n|, \qquad U_1=1\!\! 1_d,\,\,\,\,\,  \sum_{n=1}^{d^2} \omega_n=1\!\! 1_d, \en
 where $\omega_n$ is a projector since $\omega_n^2=\omega_n$, $n=1,\,\cdots,\,
 d^2$, representing a set of observables over an output parameter
 space.

\subsection{Diagrammatical rules for maximally entangled states }

\begin{figure}
\begin{center}
\epsfxsize=13.cm \epsffile{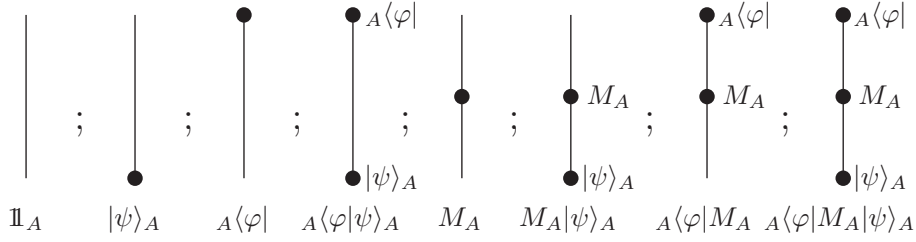} \caption{Straight lines
without or with points.} \label{fig5}
\end{center}
\end{figure}

Our diagrammatical rules assign a definite diagram to a given
algebraic expression: Every diagrammatic element is mapped to an
algebraic term. They consist of three parts: the first for our
convention; the second for straight lines and oblique lines; the
third for cups and caps.

{\bf Rule 1}. Read an algebraic expression such as an inner
product from the left to the right and draw a  diagram from
the top to the bottom.

\begin{figure}
\begin{center}
\epsfxsize=13.cm \epsffile{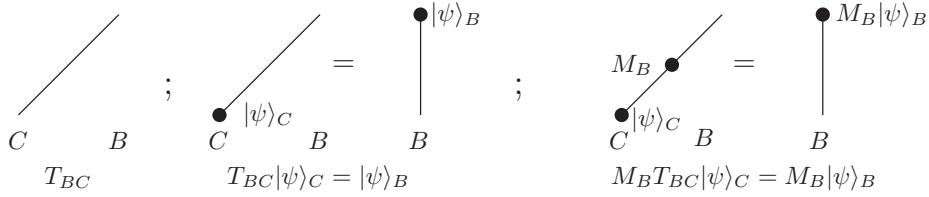} \caption{Oblique line for
transfer operator.} \label{fig6}
\end{center}
\end{figure}

{\bf Rule 2}. See Figure 5. A straight line of type $A$ denotes an
identity for the system $A$, which is a linear combination of
projectors. Straight lines of type $A$ with top or bottom boundary
solid points describe a vector $|\psi\rangle_A$, a covector
${}_A\langle \varphi|$, and an inner product ${}_A\langle
\varphi|\psi\rangle_A$ for the system $A$, respectively. Straight
lines of type $A$ with a middle solid point and top or bottom
boundary solid points describe an operator $M_A$, a covector
${}_A\langle e_i| M_A$, a vector $M_A|\psi\rangle_A$ and an inner
product ${}_A\langle \varphi | M_A |\psi\rangle_A$, respectively.
See Figure 6: An oblique line connecting the system $C$ to the
system $B$ describes the transfer operator $T_{BC}$ and its solid
points have the same interpretations as those on a straight line of
type $A$.

\begin{figure}
\begin{center}
\epsfxsize=13.cm \epsffile{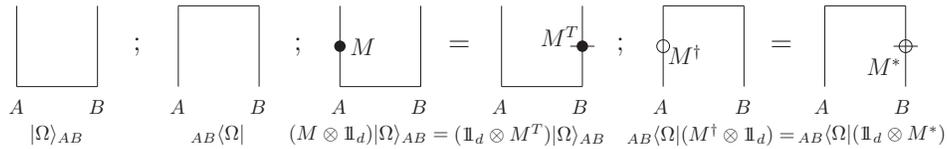} \caption{Cups and caps without
or with points.} \label{fig7}
\end{center}
\end{figure}

{\bf Rule 3}. See Figure 7. A cup denotes a maximally entangled
vector $|\Omega\rangle$ and a cap does for its dual $\langle
\Omega|$. A cup with a middle solid point at its one branch
describes the local action of an operator $M$ on $|\Omega\rangle$.
This solid point flows to the other branch and becomes a solid point
with a cross line representing $M^T$ due to (\ref{matrix}). The same
things happen for a cap except that a solid point is replaced by a
small circle to distinguish the operator $M$ from its transposed and
complex conjugation $M^\dag=(M^T)^\ast$.

 A cup and a cap can generate several kinds of configurations.
 See Figure 8. As a cup is at the top and a cap is at the bottom
 for the same composite system, such a configuration is
 assigned to a projector $|\Omega\rangle \langle \Omega |$. As a
 cap is at the top and a cup is at the bottom for the same composite
 system, this configuration describes an inner product $\langle
 \Omega|\Omega\rangle=1$ by a closed circle. As a cup is at the
 bottom for the composite system ${\cal H}_C \otimes {\cal H}_A$ and
 a cap is at the top for the composite system ${\cal H}_A\otimes {\cal H}_B$,
 the diagram is equivalent to an oblique line representing the transfer
 operator $T_{BC}$ from Charlie to Bob.

 Additionally, see Figure 8.  As a cup has a local action of the
 operator $M$ and a cap has a local action of the operator $N^\dag$,
 a resulted circle with a solid point for $M$ and a small circle for $N^\dag$
 represents a trace $\frac 1 d tr (M N^\dag)$. As conventions, we describe
 a trace of operators by a closed circle with solid points or small circles.
 We assign each cap or cup a normalization factor $\frac 1 {\sqrt d}$ and
 a circle a normalization factor $d$ according to the trace of  $1\!\! 1_d$.

\begin{figure}
\begin{center}
\epsfxsize=13.5cm \epsffile{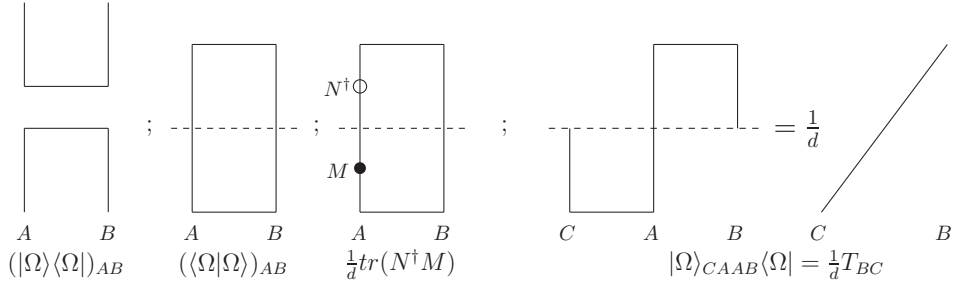} \caption{Three kinds of
combinations of a cup and a cap.} \label{fig8}
\end{center}
\end{figure}

\subsection{Description of teleportation (I): the transfer operator}

 Besides its standard description \cite{bennett} for the teleportation
 equation (\ref{tele}), the teleportation can be explained in terms of the transfer
 operator $T_{BC}$ (\ref{transfer}) which sends a quantum state
 from Charlie to Bob in the way: $T_{BC}
 |\psi\rangle_C=|\psi\rangle_B$, also see \cite{preskill}.
 In the following, we repeat the above algebraic calculation in (\ref{transfer22})
 for the transfer operator $T_{BC}$ at the diagrammatic level and then
 discuss the so called acausality problem.

From the left to the right, the inner product ${}_{CA} \langle
\Phi(U)| \Phi(V^T) \rangle_{AB}$ consists of the cap $\langle
\Omega|$, identity $1\!\! 1_d$, local unitary operators $U$ and
$V^T$, identity $1\!\! 1_d$
 and cup $|\Omega\rangle$ which are drawn from the
top to the bottom, see Figure 9. Move the local operators $U^\dag$
and $V^T$ along the line from their positions to the top boundary
point of the system $B$ and obtain the local product $(VU^\dag)_B$
of unitary operators acting on a quantum state that Bob has. The
normalization factor $\frac 1 d$ is from vanishing of a cup and a
cap. Hence the quantum teleportation can be regarded as a flow of
quantum information from Charlie to Bob.

But the operator product $\frac 1 d (VU^\dag)_B T_{BC}$ seems to
argue that the quantum measurement with the unitary operator
$U^\dag$ plays a role before the state preparation with the unitary
operator $V^T$. It is not true. Let us read Figure 9 in the way
where the $T$-axis denotes a time arrow and the $X$-axis denotes a
space distance. The quantum information flow starts from the state
preparation, goes to the quantum measurement and come backs to the
state preparation again and finally goes to the quantum measurement.
As a result, it flows from the state preparation to the quantum
measurement without violating the causality principle.

 Note that we have to impose an additional rule on how to move operators
 in our diagrammatic approach: It is forbidden for an operator to cross over
 another operator. For example, we have the operator product
 $\frac 1 d (VU^\dag)_B$ instead of $\frac 1 d (U^\dag V)_B$, see Figure 9.
 Obviously, the violation of this rule leads to the violation of the causality
 principle.

\begin{figure}
\begin{center}
\epsfxsize=13.5cm \epsffile{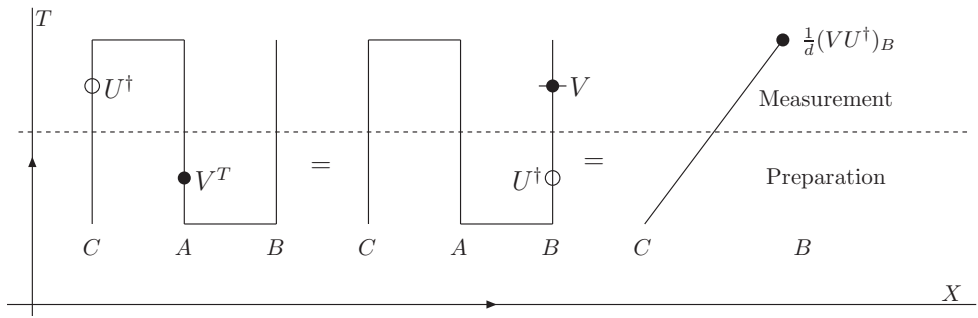} \caption{Transfer
operator and acausality problem.} \label{fig9}
\end{center}
\end{figure}

\subsection{Description of teleportation (II): quantum measurements}

Teleportation has an interpretation via quantum measurements
\cite{vaidman1, vaidman, vaidman2, eriz}. The difference from the
standard description \cite{bennett}  of the teleportation is that
the maximally entangled state $|\Omega\rangle_{AB}$ between Alice
and Bob is created in the non-local measurement \cite{vaidman1}.
Here we simply represent the quantum measurement in terms of the
projector $(|\Omega\rangle \langle \Omega|)_{AB}$.

Therefore, the teleportation is determined by two quantum
measurements: the one denoted by $(|\Omega\rangle \langle
\Omega|)_{AB}$ and the other denoted by $(|\Omega_n\rangle \langle
\Omega_n|)_{CA}$. This leads to a new formulation of the
teleportation equation,
 \eq
 \label{mtele}
 (|\Omega_n\rangle \langle \Omega_n|
 \otimes 1\!\! 1_d) (|\psi\rangle \otimes |\Omega \rangle \langle \Omega| )
 =\frac 1 d (|\Omega_n\rangle \otimes 1\!\! 1_d )
  (1\!\! 1_d \otimes (1\!\! 1_d \otimes U_n^\dag
  |\psi\rangle) \langle \Omega| ),  \en
  where the lower indices $A, B, C$ are omitted for convenience.

  \begin{figure}
\begin{center}
\epsfxsize=8.8cm \epsffile{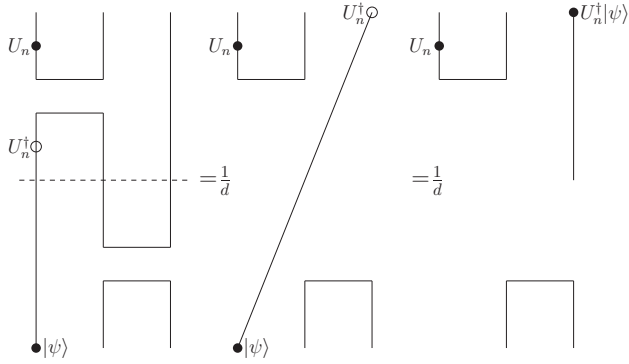} \caption{Teleportation
based on quantum measurements.} \label{fig10}
\end{center}
\end{figure}

  Read the teleportation equation (\ref{mtele}) from the left to the right
  and draw a diagram from the top to the bottom in view of our rules, i.e.,
  Figure 10. There is a natural connection between two formalisms
  (\ref{tele}) and (\ref{mtele}) of the teleportation equation. Choose all
  unitary matrices $U_n$ in the way so that they satisfy (\ref{ortho})
  and then make a summation of all independent teleportation equations like (\ref{mtele})
  to derive the version (\ref{tele}) of the teleportation equation
  in the $d$-dimension Hilbert space,
 \eq
 |\psi \rangle \otimes |\Omega\rangle = \frac 1 d \sum_{n=1}^{d^2}
  |\Omega_n\rangle \otimes U_n^\dag |\psi\rangle.
 \en

  In the case of $d=2$, the collection of unitary operators
  consists of the unit matrix $1\!\! 1_2$ and Pauli matrices $\sigma_1, \sigma_2,
  \sigma_3$. The Bell measurements are denoted by projectors
  in terms of the Bell states $|\phi^\pm\rangle$ and
  $|\psi^\pm\rangle$, and they satisfy
   \eq
 1\!\! 1_{2} =|\phi^+\rangle \langle \phi^+ |  + |\phi^-\rangle \langle
 \phi^-| +|\psi^+\rangle \langle \psi^+|  + |\psi^-\rangle \langle
 \psi^-|.
  \en
  We have the following teleportation equations, the same type as (\ref{mtele}),
 \eqa
 & & (|\phi^- \rangle \langle \phi^- | \otimes 1\!\! 1_2  )
   (|\psi\rangle \otimes |\phi^+\rangle   )
   = \frac 1 2 ( |\phi^- \rangle  \otimes  \sigma_3 |\psi\rangle ),
    \nonumber\\
 & &  (|\psi^+ \rangle \langle \psi^+ | \otimes 1\!\! 1_2  )
   (|\psi\rangle \otimes |\phi^+\rangle  )
   = \frac 1 2 ( |\psi^+ \rangle  \otimes \sigma_1 |\psi\rangle),
    \nonumber\\
 & &  (|\psi^- \rangle \langle \psi^- | \otimes 1\!\! 1_2  )
   (|\psi\rangle \otimes |\phi^+\rangle )
   = \frac 1 2 ( |\psi^- \rangle  \otimes  -i \sigma_2 |\psi\rangle), \nonumber\\
& & (|\phi^+ \rangle \langle \phi^+ | \otimes 1\!\! 1_2  )
   (|\psi\rangle \otimes |\phi^+\rangle  ) =
    \frac 1 2 (|\phi^+ \rangle  \otimes |\psi\rangle),
   \ena
 which has a summation to be the teleportation equation
 (\ref{tele}).

 The second example is on a continuous teleportation \cite{vaidman}.
 The maximally entangled vector $|\Omega^\prime\rangle$ and teleportated
 vector $|\Psi\rangle$ in the continuous case have the forms,
 \eq
 |\Omega^\prime\rangle =\int d x\,\, |x,x\rangle, \qquad |\Psi\rangle=\int d
 x\,\, \psi(x)\,\, |x\rangle,
 \en
 and the other maximally entangled state
 $|\Omega_{\alpha\beta}^\prime\rangle$ is formulated by a combined
 action of a $U(1)$ rotation with a translation $T$
 on $|\Omega^\prime\rangle$, i.e.,
 \eq
 |\Omega_{\alpha\beta}^\prime\rangle = (U_\beta \otimes T_\alpha ) |\Omega^\prime\rangle
 \equiv \int d x \exp(i\beta x) |x,x+\alpha \rangle, \qquad \alpha,
 \beta \in {\mathbb R}
 \en
  where $U_\beta|x\rangle=e^{i\beta x}|x\rangle$,
  $T_\alpha|x\rangle=|x+\alpha\rangle$  and which is a common eigenvector of
  a location operator
 $\mathbf{X}\otimes 1\!\! 1- 1\!\! 1\otimes \mathbf{X}$ and
 conjugate momentum operator
 $\mathbf{P} \otimes 1\!\! 1 +1\!\! 1 \otimes \mathbf{P}$,
 \eq
 (\mathbf{X}\otimes 1\!\! 1- 1\!\! 1\otimes \mathbf{X})|\Omega^\prime_{\alpha\beta}\rangle=
 -\alpha |\Omega^\prime_{\alpha\beta}\rangle, \qquad
 (\mathbf{P} \otimes 1\!\! 1 +1\!\! 1 \otimes \mathbf{P})|\Omega^\prime_{\alpha\beta}\rangle
 = 2 \beta|\Omega^\prime_{\alpha\beta}\rangle.
 \en
The teleportation equation of the type (\ref{mtele}) is obtained to be \eq
  (|\Omega^\prime_{\alpha\beta}\rangle  \langle\Omega^\prime_{\alpha\beta}| \otimes 1\!\!
  1 ) (|\Psi \rangle \otimes |\Omega^\prime\rangle  )
  =(|\Omega^\prime_{\alpha\beta}\rangle  \otimes 1\!\! 1 )
  (1\!\! 1 \otimes 1\!\! 1 \otimes   U_{-\beta} T_{\alpha}
  |\Psi\rangle  )
  \en
 which has a similar diagrammatic representation as Figure 10.

 Note that the continuous teleportation is a simple generalization of a discrete
 teleportation without essential conceptual changes, as is explicit in our diagrammatic
 approach.  The translation operator $T_\alpha$ is its own adjoint operator, and is
 permitted to move along the cup to the top boundary point (see Figure 10),
 although it does not behave like the matrix operator $M$ (\ref{matrix}).

\subsection{Description of teleportation (III): characteristic equations}

 In the tight teleportation and dense coding schemes \cite{werner2},
 all involved finite Hilbert spaces are $d$ dimensional and
 the classical channel distinguishes $d^2$ signals. All examples we
 treated as above belong to the tight class. In the following, we
 derive characteristic equations for all tight teleportation and
 dense coding schemes.

 Charlie has his density operator $\rho_C=(|\phi_1\rangle \langle \phi_2 |)_C$
 which denotes a quantum state to be sent to Bob. Alice and Bob share a
 maximally entangled state $\omega_{AB}$. Alice makes the Bell measurement
 $(\omega_n)_{CA}$ in the composite system between Charlie and her. As Bob gets
 a message denoted by $n$ from Alice and then applies a local unitary transformation
 $T_n$ on his observable ${\cal O}_B$, which are given by
 \eq
 T_n ({\cal O}_B ) =U_n^\dag {\cal O}_B U_n, \qquad {\cal O}_B =(|\psi_1
 \rangle \langle \psi_2 |)_B, \qquad n=1,2, \cdots d^2.
\en

 In terms of $\rho_C$, $\omega_{AB}$, $(\omega_n)_{CA}$ and
 $T_n({\cal O}_B)$, the tight teleportation scheme is summarized in
 an equation, called a characteristic equation for the teleportation,
 \eq
 \label{ttele}
 \sum_{n=1}^{d^2} tr ( (\rho\otimes \omega ) (\omega_n \otimes T_n({\cal O}) ))
 =tr(\rho {\cal O}),
 \en
where the lower indices $A, B, C$ are neglected for convenience. It
catches the aim of a successful teleportation, i.e., Charlie
performs the measurement in his system as he does in Bob's system
although they are independent of each other. This equation can be
easily proved in our diagrammatical approach. The term containing
the message $n$ is found to satisfy an equation, \eq
 tr ( ( |\phi_1\rangle \langle \phi_2| \otimes
  |\Omega\rangle \langle \Omega |) ( |\Omega_n\rangle \langle \Omega_n | \otimes
   U_n^\dag | \psi_1\rangle \langle \psi_2| U_n ) )
 = \frac 1 {d^2} tr (\rho {\cal O})
\en which is obvious in the left term of Figure 11. There  are $d^2$
distinguished messages labeled by $n$, and so we prove the tight
teleportation equation (\ref{ttele}) in a diagrammatical approach.

\begin{figure}
\begin{center}
\epsfxsize=13.5cm \epsffile{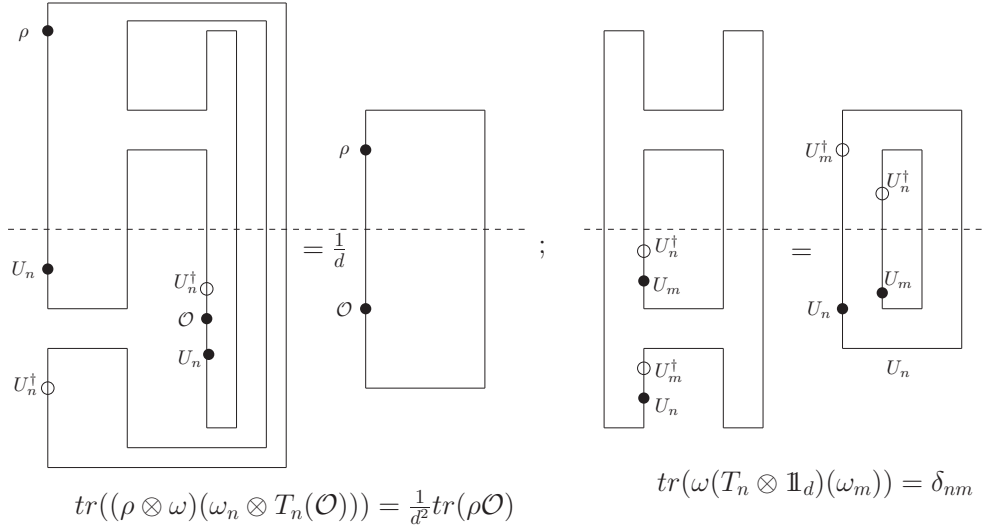} \caption{All tight
teleportation and dense coding schemes.} \label{fig11}
\end{center}
\end{figure}

A note is added for a characteristic equation for the dense coding.
As Alice and Bob share the maximally entangled state
$|\Omega\rangle_{AB}$, Alice transforms her state by the local
unitary transformation $T_n$ to encode a message $n$ and then Bob
performs the measurement on an observable $\omega_m$ of his system.
In the case of $n=m$, Bob gets the message. All the tight dense
coding schemes are concluded in the equation, \eq \label{dense}
 tr(\omega(T_n\otimes 1\!\! 1_d)(\omega_m))=\delta_{nm}
\en which can be proved in our diagrammatic approach, see the right
term of Figure 11.

\section{Generalized Temperley--Lieb configurations}

 Diagrammatical tricks involved in Figure 11 for deriving characteristic equations
 for all tight teleportation and dense coding schemes, shed us an insight that
 our diagrammatical quantum circuits have topological features to be explained by a
 mathematical structure. The diagrammatic representation for the teleportation based on
 quantum measurements, Figure 10 is a key clue for us because it is a standard configuration
 for the product $E_1 E_2$ of the Temperley--Lieb algebra as the local unitary transformation
  $U_n$ is an identity.

\begin{figure}
\begin{center}
\epsfxsize=14.5cm \epsffile{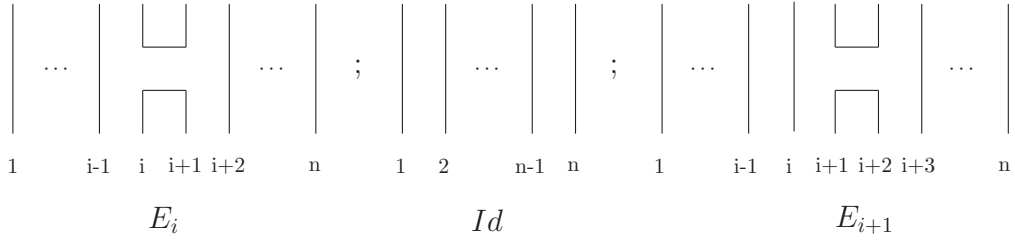} \caption{The $TL_n$
generators $E_i$, $Id$ and $E_{i+1}$.} \label{fig12}
\end{center}
\end{figure}

 We propose the Temperley--Lieb algebra under local unitary transformations to be a
 suitable algebraic structure underlying the teleportation.
 The Temperley--Lieb algebra $TL_n(\lambda)$ is generated by the identity $Id$ and
$n-1$ hermitian projectors $E_i$ satisfying \eqa \label{tl}
 E_i^2 &=&  E_i, \qquad (E_i)^\dag=E_i,\,\,\, i=1,\ldots,n-1, \nonumber\\
 E_i E_{i\pm1} E_i &=& \lambda^{-2} E_i, \qquad E_i E_j=E_j E_i, \,\,\, |i-j|>1,
  \ena
in which $\lambda$ is called the loop parameter. The diagrammatical
representation for the Temperley--Lieb algebra $TL_n(\lambda)$,
called the Brauer diagram \cite{brauer} or Kauffman diagram
\cite{kauffman12, kauffman13} in literature, consists of planar
diagrams $(n,n)$ that make connections between two rows of $n$
points. Each point in the top row is connected to another different
point in the top or bottom row by an arc, but all arcs in a diagram
have to be disjoint of each other.

See Figure 12. The identity $Id$ is represented by a diagram with
$n$ parallel vertical strings. A diagrammatical representation for
the idempotent $E_i$ is similar to the diagram for the identity $Id$
except that the  $i$th and $i+1$th top (and bottom)
 points are connected. A diagram for the idempotent $E_{i+1}$ is
 obtained in the same way. Here the configurations of cups and caps
 appear and they are formed by connecting different points in the
 same row: A cup (cap) refers to an arc between two distinct points
 at the top (bottom) row.

 Hence our diagrammatical rules take roots in the diagrammatical
 representation for the Temperley-Lieb algebra. The reason is that
 the maximally entangled state $\omega$ can form a representation of
 the $TL_n(d)$ algebra with the loop parameter $d$. For example, the $TL_3(d)$
 algebra is generated by two idempotents $E_1$ and $E_2$ given by \eq
 \label{e1e2}
 E_1=\omega\otimes Id, \qquad E_2=Id\otimes \omega,
\en which satisfy the axiom $E_1 E_2 E_1 =\frac 1 {d^2} E_1$ in the
way \eq
 E_1 E_2 E_1 |\alpha\beta\gamma\rangle =\frac 1 d \sum_{l=0}^{d-1}
 E_1 E_2 |ll\gamma\rangle \delta_{\alpha\beta}=\frac 1 {d^3}
 \sum_{n=0}^{d-1} |nn\gamma\rangle \delta_{\alpha\beta}=\frac 1
 {d^2} E_1 |\alpha\beta\gamma\rangle
 \en
 and satisfy the axiom $E_2 E_1 E_2 = \frac 1 {d^2} E_2$ via a
 similar calculation.

 \begin{figure}
\begin{center}
\epsfxsize=13.5cm \epsffile{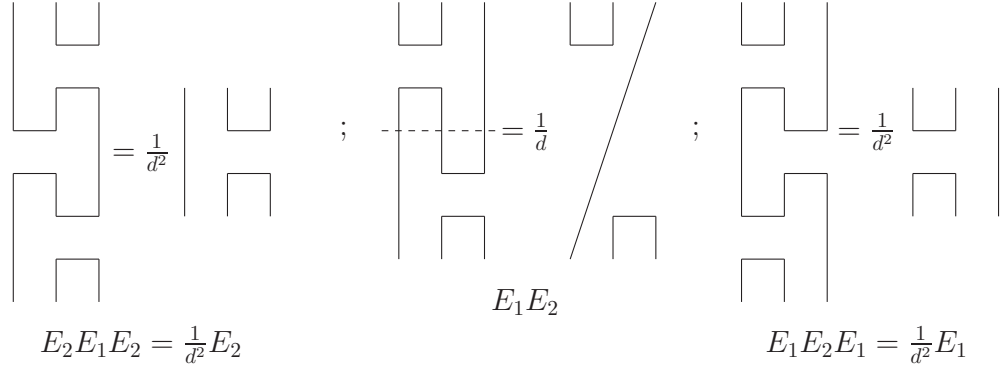} \caption{The
diagrammatical representation for the $TL_3(d)$ algebra.}
\label{fig13}
\end{center}
\end{figure}

 See Figure 13.  The product $E_i E_j$ of $E_i$ and $E_j$ is a tangle
 product obtained by attaching bottom points of $E_i$ to top
 points of $E_j$. A resulted diagram may have loops which have an
 interpretation in terms of the loop number $\lambda$. A diagram for
 the product $E_{i} E_{i+1}$ ($E_{i+1} E_{i}$) is the same as Figure 9 except
 those points representing local unitary transformations, and so it is
 called the teleportation configuration in our paper.

See Figure 13 again. We can set up the $TL_3(d)$ algebra in terms of
our diagrammatical rules by diagrammatically proving the axioms that
the idempotents $E_1$ and $E_2$ ({\ref{e1e2}}) have to satisfy. The
diagrammatical proof applies topological diagrammatical operations
by straightening configurations of cups and caps into a straight
line. Such diagrammatical tricks have been exploited in the
derivation of characteristic equations for all tight teleportation
and dense coding schemes in Figure 10. Note that each cup (cap) is
with a normalization factor $d^{-\frac 1 2}$. The normalization
factor $\frac 1 d$ in the teleportation configuration $E_1 E_2$ is
from vanishing of a cup and cap, while the normalization factor
$\frac 1 {d^2}$ in $E_1E_2E_1$ is from four vanishing cups and caps.

In terms of the density matrix $\omega_n$ (\ref{orthon}) which is a
local unitary transformation of the maximally entangled state
$\omega$, we can set up a representation of the Temperley--Lieb
algebra $TL_n(d)$. For example, the $TL_3(d)$ algebra can be
generated by
 $\tilde{E}_1$ and $\tilde{E}_2$,
 \eq \tilde{E}_1=\omega_n\otimes Id, \qquad \tilde{E}_2=Id \otimes
\omega_n, \en which are proved to satisfy the axioms of the
Temperley--Lieb algebra in our diagrammatic approach. See Figure 14.
Therefore we think that these configurations, devised for quantum
information protocols involving maximally entangled states,
projective measurements and local unitary transformations, belong to
our generalized Temperley--Lieb configurations, i.e., Temperley-Lieb
diagrams with solid points or small circles in our diagrammatic
approach. That is to say that we propose the Temperley--Lieb algebra
under local unitary transformations to underlie the quantum
teleportation phenomena.

\begin{figure}
\begin{center}
\epsfxsize=13.5cm \epsffile{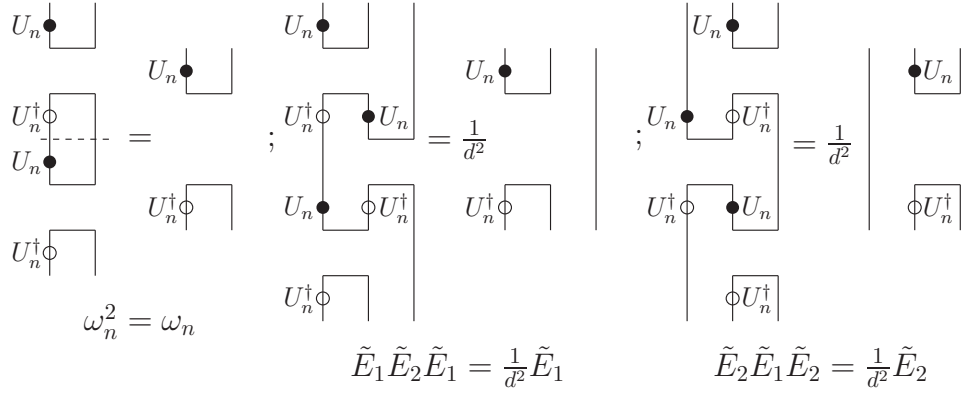} \caption{The $TL_3(d)$
algebra under local unitary transformations.} \label{fig14}
\end{center}
\end{figure}

 The teleportation configuration $E_{i} E_{i+1}$ ($E_{i+1} E_i$) is a product
 of two idempotents $E_i$ and $E_{i+1}$. Here we explain it to be a
 fundamental configuration for defining the Brauer algebra $D_n(\lambda)$
 which is an extension of the Temperley--Lieb algebra $TL_n(\lambda)$ with
 virtual crossings. The Temperley-Lieb idempotents $E_i$ and virtual crossings $v_i$
 (\ref{vbgr1}) have to satisfy the following mixed relations:
 \eqa
 \label{brauer}
 & & E_i v_i =v_i E_i =E_i, \qquad E_i v_j =v_j E_i, \qquad j\neq i\pm
 1, i=1,\cdots, n-1,
 \nonumber\\
 & & v_{i\pm1}v_i E_{i\pm1}=\lambda E_i E_{i\pm1}, \qquad
 \,\,\, E_i v_{i\pm1} v_i =\lambda E_i E_{i\pm1}.
 \ena
See Figure 15. The axiom that the teleportation configuration $E_i
E_{i+1}$ ($E_{i+1} E_i$) is equivalent to the configuration
 $v_{i+1} v_i E_{i+1}$ ($v_iv_{i+1}E_i$) informs us that the
 teleportation can be performed via two quantum gates denoted by a virtual
 crossing $v$ and the Bell measurement denoted by an idempotent $E$.
 For example, with the permutation $P$ (\ref{permutation}) as a
 virtual crossing and the  maximally entangled state $\omega$ as an
 idempotent, they  form a representation of the Brauer algebra $D_2(d)$
 with the loop parameter $d$, see \cite{ZKW} for the detail
 where $\omega$ is regarded as the partial transpose of permutation $P$.
 Hence it is interesting for quantum computing that the teleportation
 can be realized by two swap gates $P$ and the Bell measurement $\omega$.

 \begin{figure}
\begin{center}
\epsfxsize=12.5cm \epsffile{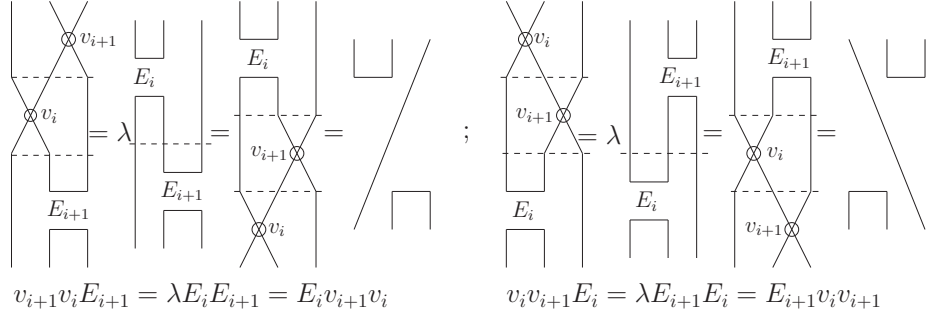} \caption{Teleportation,
teleportation swapping
 and Brauer algebra.} \label{fig15}
\end{center}
\end{figure}

 \section{Comparisons with known approaches}

 We propose the Temperley--Lieb algebra under local unitary transformations to be a
 right mathematical framework for the quantum teleportation phenomena.
 Here we compare it with two known approaches to the quantum information flow:
 the teleportation topology \cite{kauffman2, kauffman8} and strongly compact closed
 category theory \cite{coecke1}, in order to emphasize essential differences
 among them.

\subsection{Comparison with teleportation topology}

 Teleportation topology \cite{kauffman2,kauffman8} regards  the
 teleportation as a kind of topological amplitude.
  There are one to one correspondences
 between quantum amplitudes and topological amplitudes. The state
 preparation (a Dirac ket) describes a creation of two particles
 from the vacuum and has a diagrammatic representation  of a cup
 vector $|Cup\rangle$, while the measurement process (a Dirac bra) denotes
 an annihilation of two particles and is related to a cap vector
 $\langle Cap|$. The cup and cap vectors are associated with
 matrices $M$ and $N$ in the way, \eq
 |Cup\rangle = \sum_{i,j=0}^{d-1} M_{ij} |e_i\otimes e_j\rangle,
 \qquad \langle Cap | = \sum_{i,j=0}^{d-1} \langle e_i\otimes e_j |
 N_{ij}
 \en which have to satisfy a topological condition, i.e., the
 concatenation of a cup and a cap is a straight line denoted by the
 identity matrix $N_{ij} M_{jk}=\delta_{ik}$. See Figure 16.
\begin{figure}
\begin{center}
\epsfxsize=10.cm \epsffile{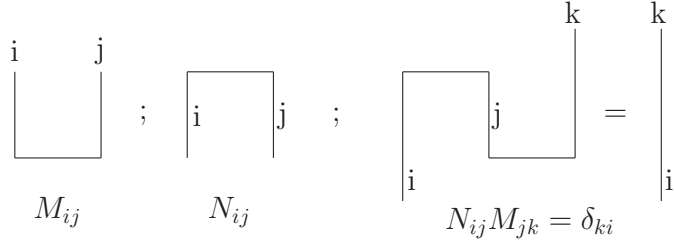} \caption{Teleportation
topology: cup, cap and topological condition.} \label{fig16}
\end{center}
\end{figure}

 However, our diagrammatic rules does not admit an interpretation of the
 teleportation topology. First, the concatenation of
 a cup and a cap is formulated via the concept of the transfer operator (\ref{transfer})
 which is not an identity required by the topological condition. Second,
 our cup and cap vectors are normalized maximally entangled vectors given by
 \eq
 |Cup\rangle =\frac 1 {\sqrt{d}} \sum_{i=0}^{d-1}  |e_i\otimes e_i\rangle,
 \qquad \langle Cap | = \frac 1 {\sqrt{d}} \sum_{i=0}^{d-1} \langle e_i\otimes e_i |
 \en which leads to a normalization factor $\frac 1 d$ to the straight
 line from the concatenation of a cup and a cap. Third, our approach underlies
 the Temperley--Lieb configurations and so involves all kinds of combinations of
 cups and caps. For example,  a projector is represented by a top cup with a bottom
 cap, as is not considered by the teleportation topology.

\subsection{Quantum information flow in the categorical approach}

We set up one to one correspondences between a bipartite vector and
a map. There are a $d_1$-dimension Hilbert space ${\cal H}_{(1)}$
and a $d_2$-dimension Hilbert space ${\cal H}_{(2)}$. A bipartite
vector $|\Phi\rangle$ has a form in terms of product bases
$|e^{(1)}_i\rangle\otimes |e^{(2)}_j\rangle $ of ${\cal H}_{(1)}
\otimes {\cal H}_{(2)}$,
 \eq
 |\Phi\rangle=\sum_{i=0}^{d_1-1}\sum_{j=0}^{d_2-1} m_{ij} |e^{(1)}_i\rangle\otimes
|e^{(2)}_j\rangle, \qquad \langle \Phi | =
\sum_{i=0}^{d_1-1}\sum_{j=0}^{d_2-1} m^\ast_{ij} \langle e^{(1)}_i
|\otimes \langle e^{(2)}_j |
 \en
 where $\langle \Phi|$ denotes a dual vector of $|\Phi\rangle$ in
 a dual product space ${\cal H}^\ast_{(1)} \otimes {\cal
 H}^\ast_{(2)}$ with the basis $\langle e^{(1)}_i |\otimes \langle e^{(2)}_j |
 $. As the product bases are fixed the bipartite vectors
 $|\Phi\rangle$ or $\langle \Phi |$ are also determined by a $d_1\times d_2$ matrix
 $M_{d_1\times d_2}=(m_{ij})$. Define two types of maps $f$ and $f^\ast$ in the
way,
 \eqa
 & & f: {\cal H}_1 \rightarrow {\cal H }^\ast_2, \qquad f(\cdot)=
 \sum_{i=0}^{d_1-1}\sum_{j=0}^{d_2-1} m_{ij} \langle
 e^{(1)}_i|\cdot\rangle \langle e^{(2)}_j |, \nonumber\\
 & & f^\ast: {\cal H}^\ast_1 \rightarrow {\cal H }_2, \qquad f^\ast(\cdot)=
 \sum_{i=0}^{d_1-1}\sum_{j=0}^{d_2-1} m_{ij}  |e^{(2)}_j \rangle \langle
 \cdot| e^{(1)}_i \rangle.
 \ena
We have the following bijective correspondences,
 \eq
 |\Phi\rangle \approx \langle \Phi | \approx M \approx f \approx
 f^\ast
 \en
which suggests that we can label a bipartite project
$|\Phi\rangle\langle \Phi|$ by the map $f$ or $f^\ast$ or matrix
$M$.

 \begin{figure}
\begin{center}
\epsfxsize=12.cm \epsffile{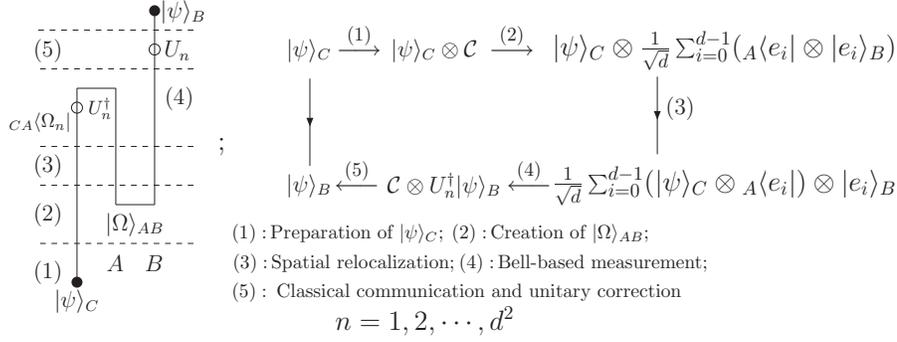} \caption{Quantum
information flow and strongly compact closed category.}
\label{fig17}
\end{center}
\end{figure}

 To transport Charlie's unknown quantum state $|\psi\rangle_C$ to Bob, the
  teleportation has to complete all the operations: preparation of
  $|\psi\rangle_C$; creation of $|\Omega\rangle_{AB}$ in Alice and
  Bob's systems; Bell-based measurement ${}_{CA}\langle \Omega_n|$ in
  Charlie and Alice's systems; classical communication between Alice
  and Bob; Bob's local unitary correction. These steps divide the
  quantum information flow into six pieces which are shown in the
  left term of Figure 17 where the third piece
  represents a process bringing Alice and Charlies' particles
  together for an entangling measurement. In the category theory,
  every step or piece is denoted by a specific map which satisfies
  the axioms of the strongly compact closed category theory. The
  crucial point is to identify a bijective correspondence between
  the Bell vector and a map from the dual Hilbert space ${\cal H}^\ast$
  to the Hilbert space $\cal H$, i.e.,
  \eq
 \frac 1 {\sqrt d}\sum_{i=0}^{d-1}
 |e_i\rangle_{A}\otimes |e_i\rangle_{B}\approx \frac 1 {\sqrt d}\sum_{i=0}^{d-1}
 {}_A\langle e_i |\otimes |e_i\rangle_{B}, \qquad {\cal H}_A\otimes
 {\cal H}_B \approx {\cal H}_A^\ast \otimes {\cal H}_B
  \en
so that the strongly compact closed category theory has a physical
realization in the form of the  quantum information flow. See the
right term of Figure 17 where the symbol
 ${\mathcal C}$ denotes the complex field and we have
 \eq
 |\psi\rangle_C\approx  |\psi\rangle_C \otimes {\mathcal C}, \qquad
 |\psi\rangle_B\approx  {\mathcal  C} \otimes |\psi\rangle_B,
 \en
and we can create a bipartite state from a complex number ${\mathcal
C}$ and also annihilate it into ${\mathcal  C}$.  Note that the
appearance of ${\cal H}^\ast\otimes {\cal H}$ is not require by the
teleportation or the quantum information flow but is imposed by the
axioms of the strongly compact closed category theory.

To sketch the quantum information flow in the form of compositions
of a series of maps which are central topics of the category theory,
we study an example in details. Set five Hilbert spaces ${\cal H}_i$
and its dual ${\cal H}^\ast_i$, $i=1,\cdots,5$ and define eight
bipartite projectors $P_\alpha=| \Phi_\alpha \rangle \langle
\Phi_\alpha |$, $\alpha=1, \cdots, 8$ in which a bipartite vector
$|\Phi_\alpha\rangle$ is an element of ${\cal H}_i \otimes {\cal
H}_{i+1}$, $i=1,\cdots, 4$. In the left diagram of Figure 18, every
box represents a bipartite projector $P_\alpha$ and a vector
$|\phi_C\rangle\in {\cal H}_1$ that Charlie owns will be transported
to Bob who is supposed to obtain a vector $|\phi_B\rangle \in {\cal
H}_5$ through the quantum information flow. The projectors $P_1$ and
$P_2$ pick up an incoming vector in ${\cal H}_2 \otimes {\cal H}_3
\otimes {\cal H}_4 \otimes {\cal H}_5 $ and the projectors $P_7$ and
$P_8$ determine an outgoing vector in $ {\cal H}_1 \otimes  {\cal
H}_2 \otimes {\cal H}_3 \otimes {\cal H}_4 $. The right diagram in
Figure 18 shows the quantum information flow from $|\phi_C\rangle$
to $|\phi_B\rangle$ in a clear way. It is drawn according to
permitted and forbidden rules \cite{coecke4}: a flow is forbidden to
go through a box from the one side to the other side, and is
forbidden to be reflected at the incoming point, and has to change
its direction from an incoming flow to an outgoing flow as it passes
through a box. Obviously, if those rules are not imposed there will
be many possible pathes from $|\phi_C\rangle$ to $|\phi_B\rangle$.

\begin{figure}
\begin{center}
\epsfxsize=12.5cm \epsffile{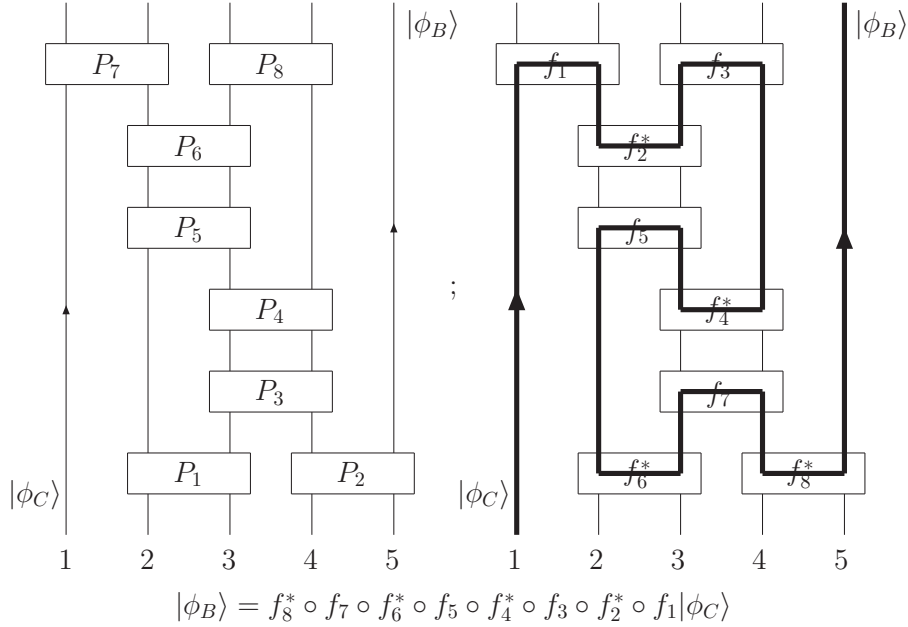} \caption{Quantum
information flow in the categorical approach.} \label{fig18}
\end{center}
\end{figure}

We now work out a formalism of the quantum information flow in the
categorical approach. Consider the projector $P_7=|\Phi_7\rangle
\langle \Phi_7 | $ and introduce the map $f_1$ to represent the
action of $\langle \Phi_7 |$, a half of $P_7$,
 \eq
f_1: {\cal H}_1 \rightarrow {\cal H}_2^\ast, \qquad
f_1|\phi_C\rangle=\langle \Phi_7 |\phi_C\rangle.
 \en
Similarly, the remaining seven boxes are labeled by the maps
$f_2^\ast$, $f_3$, $f_4^\ast$, $f_5$, $f_6^\ast$, $f_7$ and
$f_8^\ast$, respectively and so the quantum information flow is
encoded in the the form of a series of maps, \eq
 |\phi_B\rangle= f_8^\ast\circ f_7\circ f_6^\ast \circ f_5\circ f_4^\ast
  \circ f_3\circ f_2^\ast \circ f_1  |\phi_C\rangle
\en where we identify the tensor product $|\Phi\rangle\otimes 1\!\!
1_d \otimes \cdots \otimes 1\!\! 1_d$ with $|\Phi\rangle$.
Additionally, following rules of the teleportation topology
 \cite{kauffman2,kauffman8} and assigning the matrices $M, N$ to a cup
 and a cap respectively, we have the quantum information flow in the
  matrix teleportation,
 \eq
 |\phi_B\rangle= M_8\cdot N_7\cdot M_6 \cdot N_5\cdot M_4
  \cdot N_3\cdot M_2 \cdot N_1  |\phi_C\rangle.
 \en

\subsection{Comparisons with the categorical approach}

We make conceptual differences clear between our diagrammatic
teleportation approach and the categorical description for quantum
information flow. They are essential: both physical and
mathematical. In the mathematical side, we believe the braid group
and Temperley--Lieb algebra underlying the teleportation instead of
various maps in the category theory because we seek  for a real
bridge between knot theory and quantum information \cite{ZKW}. A
bipartite projector is an idempotent of the Temperley--Lieb algebra
but in the categorical approach only its half, a bipartite vector,
is exploited. In the physical side, we use concepts like the Hilbert
space, state, vector and local unitary transformation which are
basic ingredients of standard quantum mechanics described by the von
Neummann's axioms, while the categorical approach aims at setting up
a high-level approach beyond the von Neummann's axioms.

\begin{figure}
\begin{center}
\epsfxsize=9.cm \epsffile{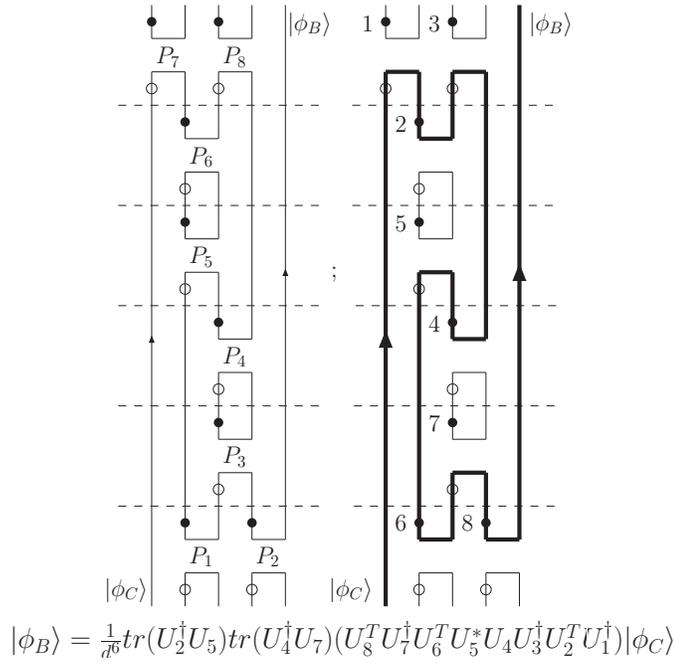} \caption{Quantum information
flow in the Temperley--Lieb configurations.} \label{fig19}
\end{center}
\end{figure}

To explain these differences in the detail, we revisit the example
in Figure 18 and redraw a diagram according to our diagrammatical
rules. In Figure 19, every projector consists of a top cup and a
bottom cap denoting maximally entangled vectors $|\Omega\rangle$ and
$\langle \Omega |$. The solid points $1,\cdots 8$ on the left
branches of cups denote local unitary transformations $U_1, \cdots,
U_8$ and small circles on the left branches of caps denote their
adjoint operators $U_1^\dag,\cdots, U_8^\dag$, respectively.
Following our rules, a quantum information flow from
$|\phi_C\rangle$ to $|\phi_B\rangle$
 is determined by the transfer operator given by
 \eq
 \label{ourflow}
|\phi_B\rangle=\frac 1 {d^6} tr(U_2^\dag U_5) tr(U_4^\dag U_7)
 (U_8^T U_7^\dag  U_6^T  U_5^\ast  U_4  U_3^\dag
  U_2^T  U_1^\dag) |\phi_C\rangle
 \en
where the normalization factor $\frac 1 {d^6}$ is contributed from
six vanishing cups and six vanishing caps, and two traces are
representations of two closed circles.

There are several remarkable things to be mentioned as Figure 18 is
compared with Figure 19. Our diagrammatical approach not only
derives the quantum information flow from $|\phi_C\rangle$ to
$|\phi_B\rangle$ in a natural way but also yields a normalization
factor from the closed circles. This normalization factor is crucial
for the quantum formation flow. For examples, setting eight local
unitary operators $U_i$ to be an identity leads to $ |\phi_B\rangle
=\frac 1 {d^4} |\phi_C\rangle $, or assuming $U_2$ and $U_5$ (or
$U_4$ and $U_7$) orthogonal to each other gives a zero vector, i.e.,
$|\phi_B\rangle=0$, no flow!

About the acausality problem which arises in the description for the
quantum information flow, we think that it is not reasonable to
argue such a question on the teleportation without considering the
whole process from the global view. See Figure 18 and Figure 19: the
quantum information flow is only one part of the diagram in our
diagrammatical approach. Note that in the categorical approach the
quantum information flow is created under the guidance of additional
permitted and forbidden rules \cite{coecke4} but in our approach it
is derived in a natural way without any imposed rules, as can be
observed in the comparison of Figure 18 with Figure 19.

 Furthermore, we can apply a bijective
correspondence between a local unitary transformation and a
bipartite vector, as is different from a choice preferred by the
categorical approach. For example, we have \eq
|\psi(U)\rangle=(U\otimes 1\!\! 1_d) |\Omega\rangle, \qquad
|\psi(U)\rangle \approx U \approx |\psi(U)\rangle \langle \psi(U)|
\en and so the Bell states (\ref{local}) are represented by
 an identity or the Pauli matrices \eq
 |\phi^+\rangle  \approx 1\!\! 1_2, \,\,\, |\phi^-\rangle  \approx \sigma_3,
 \,\,\,
 |\psi^+\rangle  \approx \sigma_1,\,\,\, |\psi^-\rangle  \approx i
 \sigma_2.
\en Hence we can regard the equation (\ref{ourflow}) as the quantum
information flow in terms of local unitary transformations.

  \section{Concluding remarks and outlooks}

In this paper, we apply the braid group, Temperley--Lieb algebra and
Brauer algebra to mathematical descriptions of the quantum
teleportation phenomena. As a remark, we explore the virtual braid
structure in the teleportation equation, and propose the braid
teleportation configuration $(b^{-1}\otimes Id)(Id\otimes b)$. They
are expected to be related to the braid statistics of anyons or
topological quantum computing \cite{wilczek}. We will make a further
study on the connection between the braid teleportation
configuration and crossed measurement \cite{vaidman}.

  In our further research \cite{yz1, yz2}\footnote{Reference [49]
  is an original version of this manuscript and it presents details of
  calculation. It has more topics which are not included here, such as
  {\em teleportation and state model,
  topological diagrammatical operations, entanglement swapping, and
  topological quantum computing}. It calls the Temperley--Lieb
  category for the collection of all the Temperley--Lieb algebra
  with physical operations like local unitary transformations.},
  we explore the braid
  teleportation configuration $(b^{-1}\otimes Id)(Id\otimes b)$ in the state model
 \cite{kauffman1} which  is devised for a braid representation of
 the Temperley--Lieb algebra and applied to a disentanglement of a
 knot or link. We find that the braid teleportation configuration consists
 of teleportation and non-teleportation terms and realize that the
 teleportation term is a basic element of the Temperley--Lieb algebra.
In addition, we study  new types of quantum algebras via the RTT
relation of the Bell matrix \cite{molin3, wang}.

We design diagrammatic rules for quantum information protocols
involving maximally entangled states, and apply them to three
different types of descriptions for the teleportation: the transfer
operator \cite{preskill}; quantum measurements \cite{vaidman1,
vaidman, vaidman2}; characteristic equations for tight teleportation
schemes \cite{werner2}. The tight teleportation scheme includes all
elements of the teleportation and unifies them from the global view
into an algebraic or diagrammatic equation,  while the other
approaches such as the standard description \cite{bennett}, the
transfer operator \cite{preskill} and quantum measurements
\cite{vaidman}, all observe the teleportation from the local point
of view.

Based on our diagrammatical representations for various descriptions
of the quantum teleportation phenomena, we propose  the
Temperley--Lieb algebra under local unitary transformations to be an
algebraic structure underlying quantum information protocols
involving maximally entangled states, projective measurements and
local unitary transformations. We find  the teleportation
configuration to be a fundamental element for defining the Brauer
algebra and obtain an equivalent realization of the teleportation in
terms of swap gates and Bell measurements.

In the extension of the present paper \cite{yz1, yz2}, we apply our
diagrammatical rules to the entanglement swapping \cite{ekert1},
quantum computing \cite{gottesman} and multi-partite entanglements.
The entanglement swapping \footnote{The entanglement swapping can be
recognized as a teleportation of a qubit which is already entangled
with another qubit.} yields an entangled state between two systems
which never met before and have no physical interactions, and it has
a characteristic equation which can be derived in our diagrammatical
approach. In the quantum computing which chooses the teleportation
as a primitive element, we show the Temperley--Lieb configurations
for a unitary braid gate, swap gate and CNOT gate. Additionally, we
discuss how to represent multipartite entangled states in our
diagrammatic approach.

 In a general sense, Bell measurements and local unitary transformations
 are crucial points for an application of our diagrammatic rules.
 We expect our diagrammatic approach to be exploited in topics
 such as Bell inequalities, quantum cryptography, and so on. To
 develop our diagrammatical rules, we refer to the work on atemporal
 diagrams for quantum circuits \cite{griffths} where a system of
 diagrams is introduced for the representation of various elements
 of a quantum circuit in a form which makes no reference to time.
 Note that our diagrammatical rules are devised for uncovering
 topological features in quantum circuits but an atemporal diagram
 is a kind of diagrammatical notation for a quantum circuit.

 In our further research, we will continue to explore
 applications of mathematical structures involved in this paper.
 We expect the categorical theory to play important roles in
 the study of the quantum information phenomena in view of
 \cite{coecke5,duncan, baez, selinger,kauffman14}. We think that
 the Temperley--Lieb algebra under local unitary transformations,
 is an interesting mathematical subject, because it contains abundant
 mathematical objects such as  braids and provides a kind of realization
 of quantum computing. As a virtual braid representation can be
 obtained via the Brauer algebra (i.e., the virtual Temperley--Lieb
 algebra \cite{ZKW, ZKW1}),  we need set up connections
 between the virtual braid teleportation configuration and virtual
 Temperley--Lieb configuration.

\section*{Acknowledgements}

 I am indebted to X.Y. Li for his constant encouragements and supports,
 and grateful to L.H. Kauffman  for his helpful comments on the
 teleportation topology. I thank M.L. Ge for helpful discussions on
 the rules for oblique lines, thank L. Vaidman for his helpful email correspondences
 on the crossed measurement, and thank R.F. Werner for his helpful comments
 on tight teleportation schemes and his email correspondence on the project of
 setting up a bridge between knot theory and quantum information.
 This work is in part supported by NSFC--10447134 and SRF for ROCS, SEM.
 I thank two referees of my manuscript for their critical comments
 on the presentation style of results and their helpful interpretations
 on the quantum teleportantion. The second, third and fourth footnotes are taken
 from referees' reports.

 \end{document}